\documentclass[twocolumn,pre,aps]{revtex4}
\usepackage{graphics,graphicx,epsfig}
\usepackage{epsf,epstopdf,wrapfig}
\usepackage{amssymb,amsfonts,amsmath}
\usepackage{graphicx,pstricks,epsfig}
\usepackage{ifthen}
\include{epsf}

\newcommand{\<}{\langle}
\renewcommand{\>}{\rangle}
\newcommand{\refappendix}{the appendix}
\newcommand{\refappendixfig}{\ref{panel1}}

\newcommand{\beq}{\begin{equation}}
\newcommand{\eeq}{\end{equation}}
\newcommand{\bea}{\begin{eqnarray}}
\newcommand{\eea}{\end{eqnarray}}

\newcommand{\bs}{ \mbox{\boldmath$\sigma$}}

\begin{document}

\title{Maximum entropy models for antibody diversity}

\author{Thierry Mora\footnote{These authors contributed equally.}$^{1}$, Aleksandra M. Walczak$^{*1,2}$, William Bialek$^{1,2}$ and Curtis G. Callan, Jr.$^{1,2}$}
\affiliation{$^1$Joseph Henry Laboratories of Physics, Lewis--Sigler Institute for Integrative Genomics, and $^2$Princeton Center for Theoretical Science,
Princeton University, Princeton, New Jersey 08544 USA}

\date{\today}

\begin{abstract}
Recognition of pathogens relies on families of  proteins showing great diversity.
Here we construct maximum entropy models of the sequence repertoire, building on recent experiments that provide a nearly exhaustive sampling of the IgM sequences in zebrafish.  These models are based solely on pairwise correlations between residue positions, but correctly capture the higher order statistical properties of the repertoire.  Exploiting the interpretation of these models as statistical physics problems, we make several predictions for the collective properties of the sequence ensemble:  the distribution of sequences obeys Zipf's law, the repertoire  decomposes into several clusters, and there is a massive restriction of diversity due to the correlations.  These predictions are completely inconsistent with models in which amino acid substitutions are made independently at each site, and are in good agreement with the data.  Our results suggest that antibody diversity is not limited by the sequences encoded in the genome, and may reflect rapid adaptation to antigenic challenges. This approach  should be applicable to the study of the global properties of other protein families.
\end{abstract}

\maketitle

\section{Introduction}

The number of possible amino acid sequences exceeds the number of individual protein molecules that have ever been synthesized.  As a result, the limited set of sequences that we see today carries a signature of evolutionary history \cite{Pal}.  But not all of the limitations are historical---randomly chosen sequences will not fold into stable, compact structures \cite{Branden,Cordes:1996p5095}, and carrying out specific functions places yet more requirements on the sequence.  Regardless of the balance between historical and functional constraints,  the stochastic nature of evolutionary change means that the sequences we observe should be thought of as being drawn out of a probability distribution.   The goal of this paper is to construct an approximation to this distribution, using a limited but biologically important example, the problem of antibody diversity.

The ensemble of {\em all} proteins is daunting, so most work focuses on particular families of proteins.  The most tractable examples are those in which the relevant segments of the proteins are short, and experiments provide many independent samples of sequences from the family.  For a family of small proteins that mediate protein--protein interactions, methods were developed to generate new sequences that are consistent with the patterns of single site substitutions and correlations between substitutions at pairs of sites; remarkably, most of these new sequences fold into functional structures  \cite{Socolich:2005p1730,Russ:2005p1728}.  Although this work did not lead to an explicit construction of the underlying probability distribution, the implicit model is equivalent to a maximum entropy model that captures pairwise correlations but ignores higher order interactions \cite{Bialek:2007p3854}, and thus connects to other efforts to describe biological networks with simplified models \cite{dhadialla,Schneidman:2006p1273,seno,tangetal,preprint,volkov}.  Maximum entropy methods have since been used  to look at protein--protein interactions in bacterial signaling \cite{Weigt:2009p3341}, and at the serine proteases \cite{Halabi:2009p3278}.

A key feature of the maximum entropy approach is its intimate connection to statistical mechanics \cite{Jaynes:1957p4009,Jaynes:1957p4011}.   Maximum entropy models predict the underlying probabilities in the form of a Boltzmann distribution, thus assigning an effective energy to every amino acid sequence in our ensemble.  Natural questions about this statistical mechanics problem have clear biological correlates:  What is the entropy in sequence space, or equivalently the allowed diversity of functional proteins?  Does the energy landscape break up into multiple valleys, corresponding to clusters of closely related proteins?  Are the barriers between these valleys large, so that different clusters are isolated, or are there paths that can smoothly mutate one class of sequences into another?  Are the interactions among substitutions at different sites strong or weak?  Is is possible that these interactions are tuned to some special values, perhaps analogous to critical points in statistical mechanics?  Here we approach these problems in the context of antibody diversity.  

For antibodies, sequence diversity has  a direct biological function, setting the range of antigenic challenges to which the organism can respond.  Classical  work has emphasized the combinatorial diversity generated by piecing together different segments of the antibody molecule, each of which is encoded in the genome \cite{Hozumi:1976p5097}.
Very recently it has become possible to provide the sequences of essentially every single antibody molecule in individual organisms \cite{Weinstein:2009p1566}, and this explosion of data  invites us to look more closely at the diversity within the combined segments, beyond that represented in the genome itself.  As we will see, for the zebrafish studied in Ref \cite{Weinstein:2009p1566}, this non--genomic diversity is substantial, and concentrates in short segments of the molecule, the D regions of these molecules.     This combination of focus on short sequences and a nearly complete sampling of the relevant ensemble provides a unique opportunity to address the theoretical questions outlined above.

\section{Defining the problem}

All jawed vertebrates are endowed with an adaptative immune system that responds to and `remembers' a wide range of challenges from the environment.
One major component of the immune system are the B cells, each of which expresses multiple copies of a single antibody molecule on its surface. Binding to these molecules is the fundamental step by which the system recognizes an antigen, and hence the diversity of these molecules defines the range of pathogens to which the organism can respond effectively \cite{Janeway}.   During the development of B cells, the genome is modified by recombination to encode a single antibody sequence assembled from three pieces termed V, D and J.  In the zebrafish \cite{Lieschke:2009p5098}, there are 39 choices for the V region, 5 for D, and 5 for J, for a total of 975 possible VDJ combinations or classes. During recombination, nongenomic nucleotides are randomly added and others are removed at the VD and DJ junctions, generating what is called junctional diversity.  Further, during the lifetime of the organism, the antibody sequences encoded in proliferating B cells undergo somatic hypermutation.  Finally, B cells that successfully bind pathogens proliferate, while B cells that are not used are eliminated.  As a result, the expressed repertoire of antibodies is a complex combination of VDJ class, phylogenic history and pathogen environment.

\begin{figure}
\noindent\includegraphics[width=\linewidth]{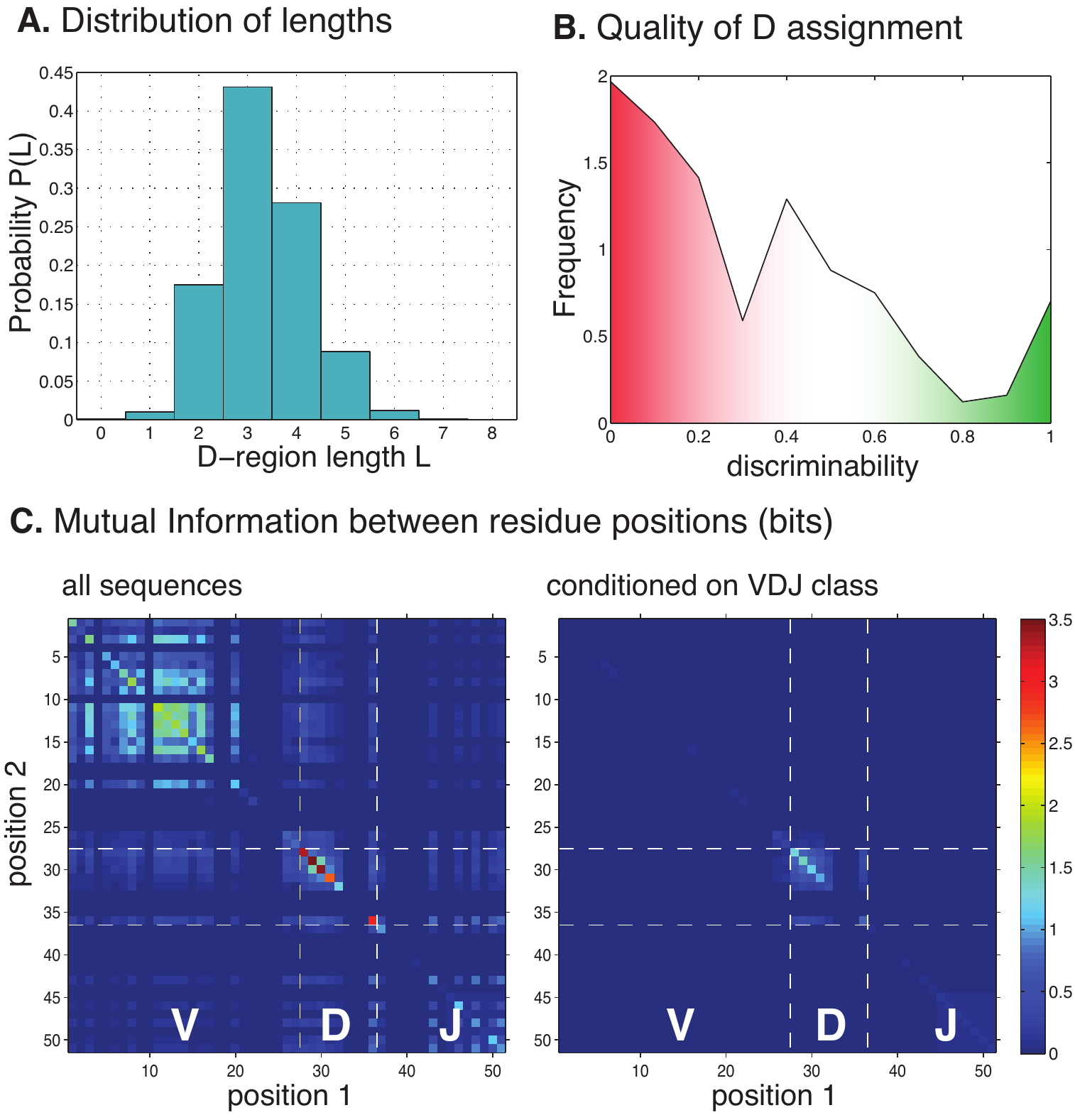}
\caption{{\bf Antibody diversity concentrates in the D region} (data from fish A). {\bf A} Length distribution of the D region aminoacid sequences.  {\bf B.}  Quality of the assignment of the D region to a genomic template, measured by the difference between the alignment scores of the first and second best matches (normalized by the best attainable score difference). {\bf C.} Left: mutual information between residue positions across all VDJ classes. Right: mean mutual information within each VDJ class. Variability only remains in the D region.
\label{panel1}}
\end{figure}

The experiments of Ref \cite{Weinstein:2009p1566} give us a snapshot of the complete antibody repertoire in each of fourteen zebrafish, labelled A through N.  More precisely, these experiments extracted the mRNA for the complementarity determining region 3 (CDR3) of the heavy chain of IgM molecules, reverse transcribed, amplified, and then sequenced the resulting cDNA using high throughput methods.  It will be important in our analysis that the amplification step has biases, and so all averages over the distribution of sequences must be re--weighted by a primer dependent amplification, as discussed in \cite{Weinstein:2009p1566} (see \refappendix).  Each fish yielded from 28,000 to 112,000 sequence reads of $\sim 200$ nucleotides covering the last $~$90 nucleotides of V, and all of D and J. 

The V and J segments of all the sequences are easily recognized by aligning with the genome, discarding a small fraction of sequences with stop codons or frame mismatches.  The situation for D regions is more subtle, and so we define the D region to be all the residues that lie between the identifiable parts of the V and J segments, as explained more fully in {\refappendix}.

We find that the D region is much more diverse than expected from its genomic origin, and concentrates most of the nongenomic diversity, as illustrated in Fig.~{\refappendixfig}. Most obviously, in the genome D regions range from 11 to 14 nucleotides, while in the sampled sequences the D region range from 1 to 6 amino acids (3 to 18 nucleotides; Fig {\refappendixfig}A).  If we try to match each sequence to one of the genomic sequences, the quality of these assignments typically is quite poor (Fig {\refappendixfig}B).
By using mutual information between residue positions as a measure of variability within VDJ classes (see {\refappendix}), we find that residues in the D region are both variable and correlated even within a given D class, whereas the V and J regions show very little diversity within their classes  (Fig {\refappendixfig}C).  
Junctional diversity, somatic hypermutations or other mechanisms may be the source of this nongenomic D variability,  and could explain the poor quality of the D assignments.  Independent of the mechanism, these results suggest that, in trying to define the distribution of  sequences represented in the system, we should focus our attention on the D region.

To be precise, we describe each observed D region sequence as $\bs=(\sigma_1,\, \sigma_2,\, \cdots,\, \sigma_{L})$, where $L$ is the length of the sequence.  At each site along the sequence, $\sigma_{\rm i}$ can take on twenty different values, corresponding to the twenty possible amino acids ($\sigma_{\rm i}$ = Ala, Arg, Asn, \ldots).  We would like to know the probability $P(\bs )$ that any particular sequence will be found in the antibody repertoire of each individual.  The difficulty is that there are $\sim (20)^{L_{\rm max}}$ possible sequences, where $L_{\rm max} = 8$ is the maximum length of the D region; in principle each sequence can occur with a different probability, and hence the number of possible sequences is also the number of parameters required to specify the distribution.  This number, $\sim 2.5\times 10^{10}$, is much larger than the number of independent measurements that we can make, and perhaps even larger than the number of B cells in the entire zebrafish at any one moment.  How, then, can we make progress?

\begin{figure}
\noindent\includegraphics[width=\linewidth]{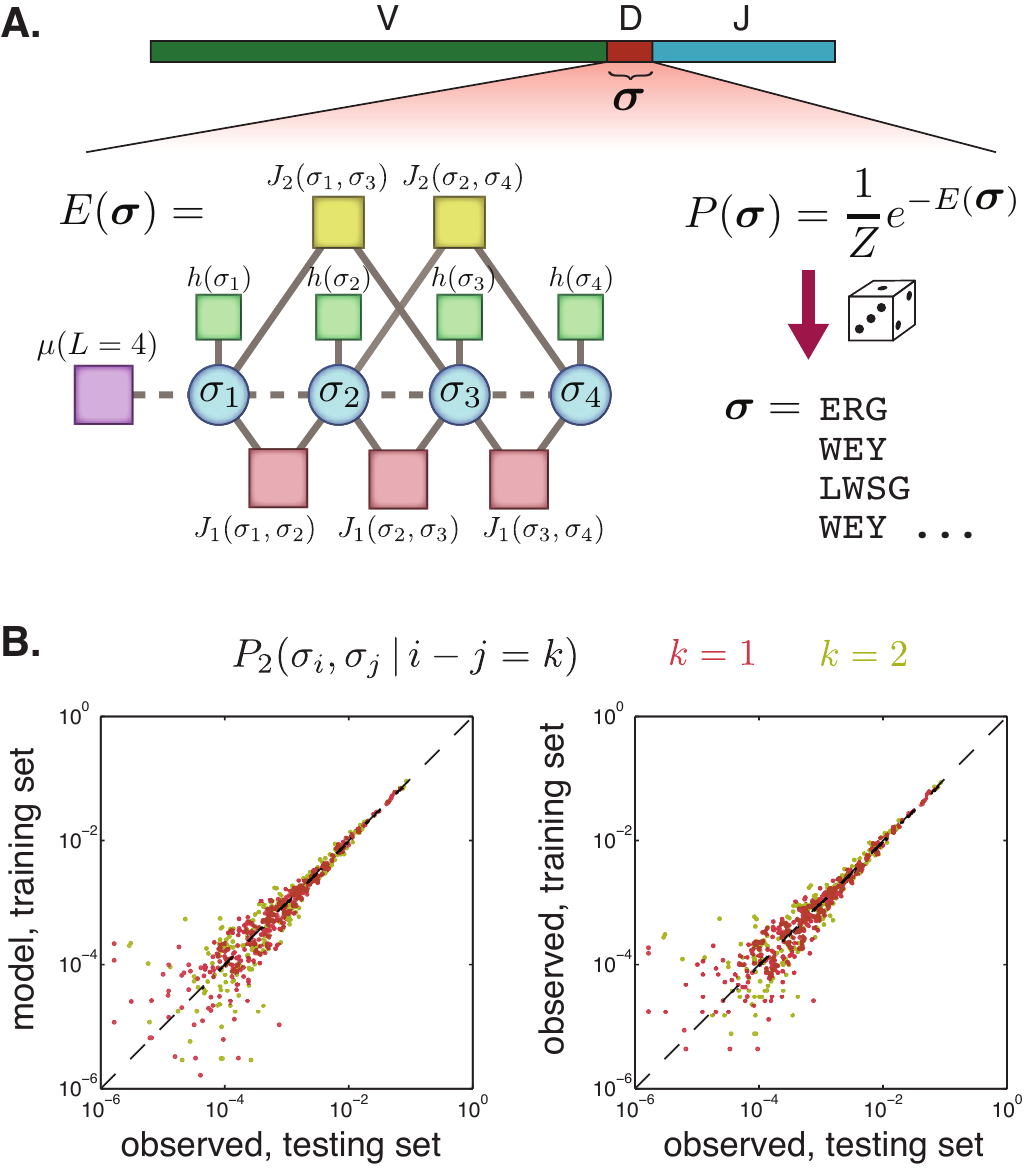}
\caption{{\bf Maximum entropy model. A. } The model of the D region is viewed as a system of interacting residues $(\sigma_1,\ldots,\sigma_L)$ in thermal equilibrium, schematized here by its interaction network for $K=2$.
To each sequence $\bs$ is associated an energy $E(\bs)$, Eq (\ref{energy}). Then the sequences of the repertoires are drawn at random from the Boltzmann distribution , Eq (\ref{boltzmann}).
{\bf B. } Pairwise frequencies of nearest-- (red) and second--neighbor (yellow) residues. Left: comparison between the model prediction, where the model was fitted with the training data, and the testing data. Right: direct comparison between the training data and the testing data. The scatter is of the same magnitude, showing that the model is as precise as the data allow.
\label{panel2}}
\end{figure}

\section{Maximum entropy models of the D region}

While experiments cannot characterize the entire distribution $P(\bs )$, it is possible to make reliable measurements of many averages over this distribution.  For example, we can characterize the probability that any single amino acid appears in the sequence, $P_1(\sigma )$.  Further, we can characterize the probability that two particular amino acids appear separated by a distance $\rm k$ along the sequence, $P_2(\sigma , \sigma '; {\rm k})$, and we can do this for nearest neighbors (${\rm k} =1$), next  nearest neighbors (${\rm k} =2$),  and so on.  Notice that these quantities do not refer to specific sites along the sequence, but rather to pairs of sites separated by given distances; in this way we can analyze sequences that have variable lengths and are difficult to align, as observed for the D regions.  We could continue along this line, characterizing the probability of occurrence of triplets, quartets, etc., but at some point we will run out of data.    
 
The central idea of maximum entropy models is to take some limited set of averages seriously as a characterization of the system, and then build the least structured model for the distribution $P(\bs )$ that is consistent with these data
\cite{Jaynes:1957p4009,Jaynes:1957p4011}.  Formally, minimizing structure means maximizing the entropy 
\begin{equation}\label{entropy}
S[P]=-\sum_{\bs} P (\bs) \log_2 \left[P (\bs)\right].
\end{equation}
Here we will find the maximum entropy distribution consistent with the single residue frequencies, $P_1 (\sigma)$, with the pairwise distributions of amino acids along the sequence, $P_2(\sigma , \sigma '; {\rm k})$, and with the observed distribution of lengths of the D region, $P(L)$.  Finding this model distribution, which we denote $P^{({\rm m})}$, involves solving an optimization problem (maximize $S$) subject to constraints (the observed distributions).    Because of the connection between maximum entropy distributions and statistical mechanics, the form of the solution is well known.  

We can write $P^{({\rm m})}$ in the form of the Boltzmann distribution, as if the sequences represented the state of a physical system in thermal equilibrium:
\begin{equation}
P^{({\rm m})} = {1\over Z} \exp\left[ - E(\bs )\right],
\label{boltzmann}
\end{equation}
where the effective energy of each sequence is
\beq\label{energy}
E(\bs)=-\mu(L)-\sum_{{\rm i}=1}^L h(\sigma_{\rm i})-
\sum_{{\rm k}=1}^{K} \sum_{\substack{{\rm i,j}\\{\rm i}-{\rm j}={\rm k}}} J_{\rm k}(\sigma_{\rm i}, \sigma_{\rm j}),
\eeq
To complete the analogy to thermodynamics, we should think of the temperature as being such that $k_B T = 1$.  Then $\mu (L)$ acts like a chemical potential for adding residues, $h(\sigma )$ is a biasing field that prefers some amino acids over others, and the couplings $J_{\rm k}$ describe the interactions between amino acids at different sites, reaching across a range $K$, as schematized in Fig \ref{panel2}A.
The $h$'s, $J$'s  and $\mu$'s must be  chosen such that $P^{\rm (m)} (L)$, $P^{\rm (m)}_1$ and $P^{\rm (m)}_2$ agree with the data.

Calculating $P^{\rm (m)} (L)$, $P^{\rm (m)}_1$ and $P^{\rm (m)}_2$ from the full distribution $P^{\rm (m)}(\bs )$  is hard in general, and the inverse problem of inferring the model parameters from these observables is clearly not easier.
We solve the inverse problem by combining Monte Carlo simulations with gradient descent (see {\refappendix}). The number of parameters can be fairly large, $399K+19+L_{\rm max} \sim 10^3$, although vastly smaller than the number of possible parameters $(20)^{L_{\rm max}}$.
To test the validity of our method and control for overfitting, we learned the maximum entropy distribution from only one half of the sequences (training set). Then, the model predictions  were compared to the second half of the data (testing set). We solved the inverse problem and tested our solution for all 14 fish and for different interaction ranges $K=1,\, 2,\, 3,\, 4$.
Our results showed excellent agreement with the data, as illustrated in Fig.~\ref{panel2}B for the pairwise frequencies in fish A ($K=2$).

\begin{figure}
\noindent\includegraphics[width=\linewidth]{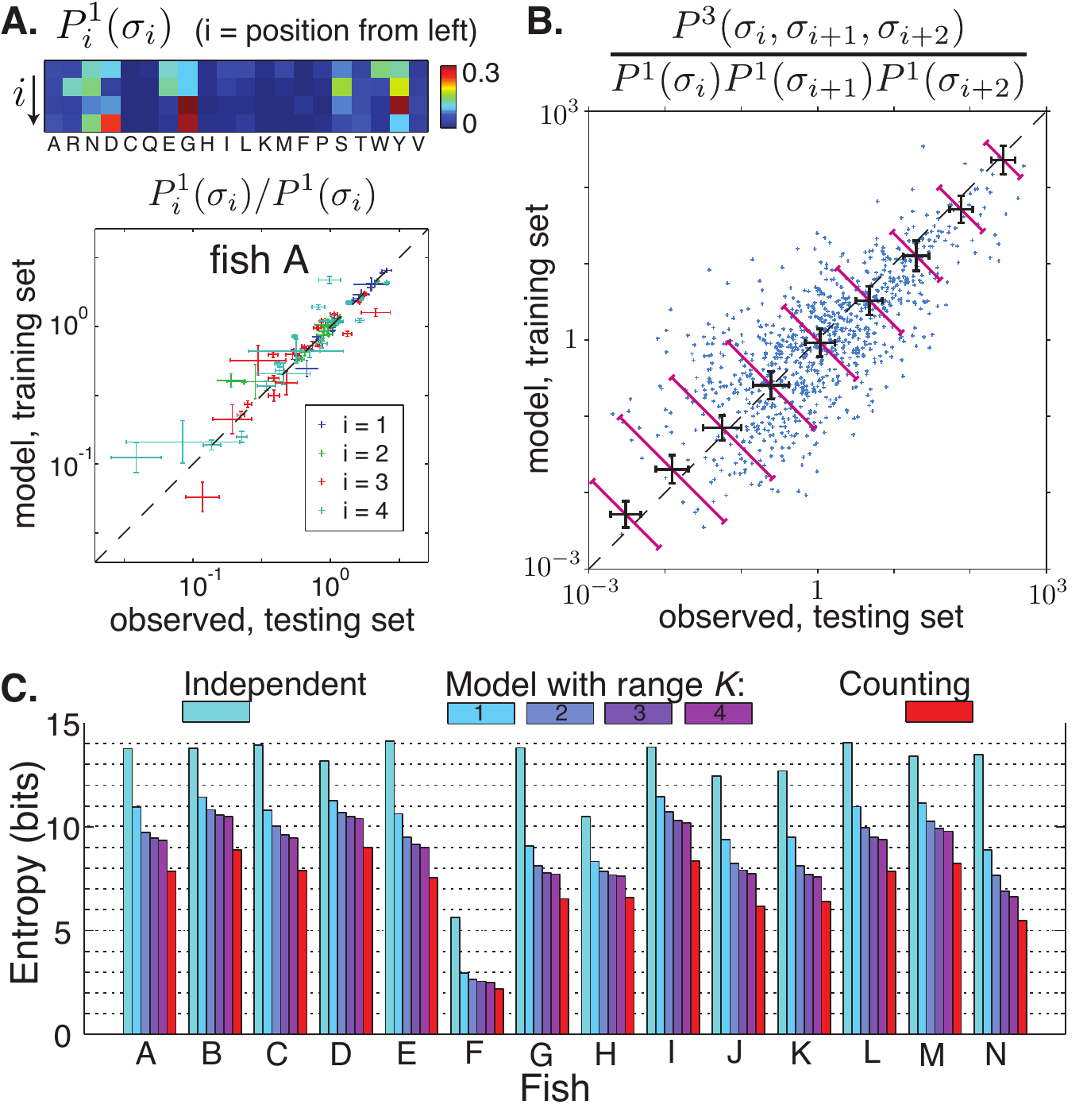}
\caption{{\bf Local observables and the entropy are well captured by the model.} {\bf A.} Position-dependent aminoacid frequency. Top, frequency as a function of position $i=1,\ldots,4$ from the left end of the sequence. Bottom: comparison between model and data of position-dependent frequencies, normalized by the prediction of the independent model. Error bars are obtained as the standard deviation over many choices of partition between training and testing sets.
 {\bf B.} Comparison of triplet frequencies of contiguous aminoacids, normalized by the prediction of the independent model. The small crosses illustrate one choice of the training/testing partition. The black error bars represent the average measurement error made on a triplet frequency at that frequency value, obtained as the standard deviation over many choice of the training/testing partition. The diagonal error bars show the average error between model and data.
{\bf C.} Entropy of all fish: from frequency counting, from the independent model, and from the maximum entropy model with range $K=1,\ldots,4$.
\label{panel3}}
\end{figure}

\section{Testing and exploring the model}

The maximum entropy model is the least structured model consistent with the observed pairwise correlations among amino acids, but of course there is no guarantee that Nature is described by this minimal model.  To test the model, we  look systematically at its predictions for measurable quantities that are not already used in determining the model parameters.  If we can convince ourselves that these predictions are at least approximately correct, we can   take the model more seriously and ask what it tells us about the nature of antibody diversity.

\subsection{Local biases}

The model we have constructed does not incorporate any site specificity---interactions between amino acids  depend on the distance between them but not on their absolute location along the sequence [Eq (\ref{energy})].  But, since amino acids at the start  or end of the sequence have only half the number of neighbors that are available to sites in the middle of the sequence, the model predicts `end effects' which will be manifest as position specific biases in amino acid composition.    As shown in Fig \ref{panel3}A, these predicted biases can be large, so that the probability of finding particular amino acids at specific sites, $P_{\rm i}^{1}(\sigma )$,  can vary by more than two orders of magnitude.  These predictions are in very good agreement with the data.  We emphasize once again that these predictions of site specific substitution patterns are obtained from a model that has no explicit site specific information, and which is learned from an ensemble of sequences that have not been aligned.  In a similar spirit, we find good agreement between the predicted and observed probabilities of contiguous amino acid triplets (Fig.~\ref{panel3}B), even though the model has no explicit three site interactions.

\subsection{Zipf's law}

\begin{figure}
\begin{center}
\noindent\includegraphics[width=.7\linewidth]{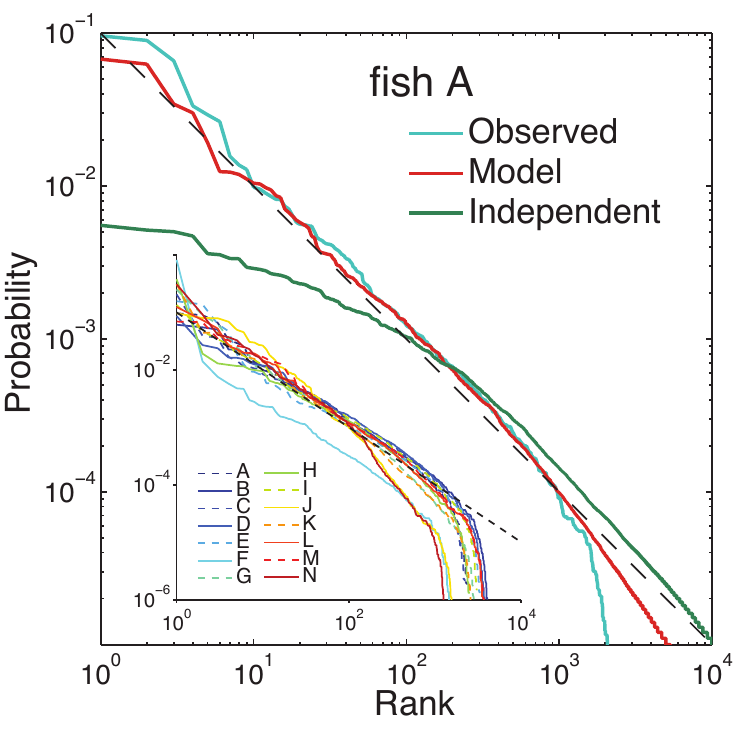}
\end{center}
\caption{{\bf The distribution of D regions obeys Zipf's law.} Probability of D region sequences as a function of their rank in fish A, as observed from frequency counting (blue line), and as predicted by the independent (green line) and the maximum entropy model with $K=2$ (red line). The dashed line has slope $-1$. Inset: the same for all fish, from frequency counting.
\label{zipf}}
\end{figure}

The space of possible sequences is so large that we cannot  test the predictions for the distribution $P(\bs )$ directly.  Still, we can get a global view of the distribution through a Zipf plot, in which we put the observed sequences in order based on their frequency of occurrence, and plot probability $P$ vs. rank $r$, as in Fig \ref{zipf}.   We see that both the data and the predictions of the model are very close to obeying Zipf's law, $P \propto 1/r$ \cite{Zipf,Newman:2005p4243}, and the data and model agree very well with one another.  The same pattern is observed in all fish, although the ranking of particular sequences varies.  The dynamic range over which we can observe Zipf's law is limited by the number of independent sequences that are read in the experiments, but the model predicts that this behavior should continue even if this number were extended by an order of magnitude.

Zipf's law first attracted attention in the context of language \cite{Zipf}, and many models have been proposed for the origin of this behavior.  Even before Zipf's work, it was known that some growth processes with mutations can generate Zipfian distributions \cite{Yule:1925p3020,Newman:2005p4243}.  Since we have built  a model out of measured pairwise correlations, with strong analogies to statistical mechanics, we emphasize that Zipf's law reflects the proximity of a critical point in the strength of the underlying interactions.  The rank of a state $\bs$ is determined by the number of states with higher probability, or lower energy in Eq (\ref{boltzmann}).  
But counting the number of states is equivalent to measuring the (microcanonical) entropy, and then Zipf's law is the statement that the entropy grows linearly with the energy, with slope one (see  {\refappendix}).   
This locally linear relation between energy and entropy is characteristic of thermodynamic systems at a critical point \cite{Huang}, and could not emerge from a system of noninteracting units, or even from an interacting system with slightly weaker or stronger correlations.  Thus, the strength of correlations that we see in the real sequences corresponds to interactions with a critical strength, restricting the set of allowed sequences substantially, but not forcing the system to `freeze' into a small set of possibilities.  

\subsection{Entropy}

The fundamental quantity in a maximum entropy construction is the entropy $S$ itself.  Entropy measures the diversity in sequence space, and hence is also a fundamental quantity from a biological point of view.  If we imagine that sequences are constructed by choosing amino acids at random, then the entropy could be as large as $\log_2(20)\,{\rm bits}$ per residue, or a total of 
$\sim 15\,{\rm bits}$ for the average length D region.  For almost all fish (F is an exception, and is excluded from further analyses), the observed biases in the use of the different amino acids do not reduce this very much; that is, if we choose amino acids independently at every site but with the observed frequencies,
\begin{equation}
P_{\rm ind}(\bs)\equiv P(L)\prod_{i=1}^L P_1(\sigma_i) ,
\label{Pind}
\end{equation}
then the entropy $S[P_{\rm ind}]$ of this independent model is nearly $\log_2(20)\,{\rm bits}$ per residue.
We can think of the maximum entropy model as part of a hierarchy, in which the entropy is reduced every time we take account of additional correlations \cite{Schneidman:2003p4135}.  As shown in Fig \ref{panel3}C, the entropy is reduced significantly as we take account of correlations between neighboring amino acids, corresponding to $K=1$ in Eq (\ref{energy}).  It is reduced further when we include next--nearest neighbors ($K=2$), and the reduction seems to plateau as we include more distant neighbors, $K=3,\,4$.    Including all of these pairwise correlations pushes the total entropy well below 10 bits for all fish, so that out of tens of thousands of possible sequences, most of the distribution is concentrated in only a few hundred ($\sim 2^S$) sequences, and this is consistent with what we observe in the Zipf plots (Fig \ref{zipf}).  This restriction of sequence space is even more dramatic when we realize that, given the maximum length of the D regions, there really are tens of millions of possible sequences.  

The difference between the entropy of the independent model and the true entropy, $I = S[P_{\rm ind}] - S[P]$, measures the overall strength of correlations in the system, and is called the multi--information.  The maximum entropy model predicts a value for $I^{\rm (m)} = S[P_{\rm ind}] - S[P^{\rm (m)}]$ which must be smaller than $I$, and the ratio $I^{\rm (m)}/I$ meausres the fraction of the correlated structure that we capture in our model.  The difficulty is that because sequence space is large, estimating the entropy $S[P]$ is difficult.  Methods are available, however, which allow us to estimate $S[P]$ even when we don't have enough samples to accurately estimate $P(\bs )$ itself, as explained in {\refappendix} and \cite{Strong}.  Using these methods, we find, as shown in Fig \ref{panel3}C, $I^{\rm (m)}/I$ in the range from $0.67$ to $0.91$ across the different fish.
Thus our maximum entropy model, based only on pairwise correlations, captures between two thirds and ninety percent of all the correlated structure in the distribution of sequences.  

\begin{figure}
\begin{center}
\noindent\includegraphics[width=.8\linewidth]{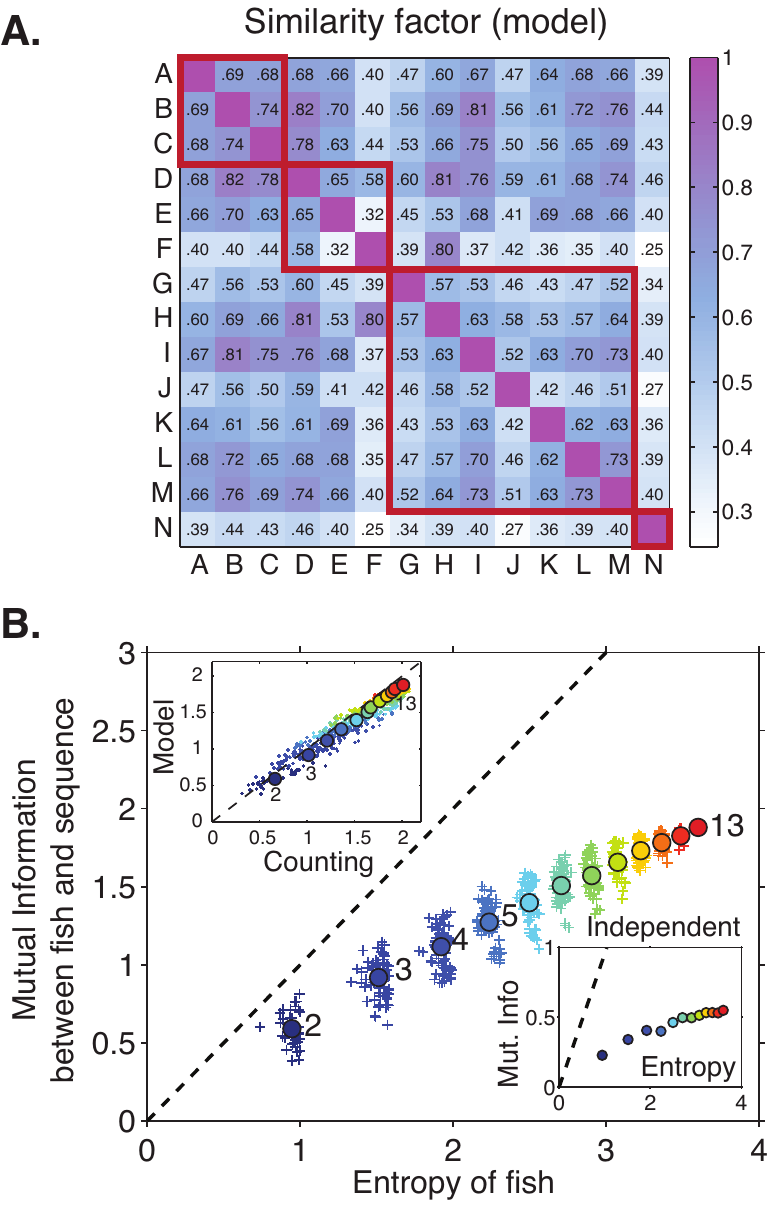}
\end{center}
\caption{{\bf Fish repertoires overlap yet are specific. A.} Table of similarity factors between all pairs of fish. The similarily factor is defined as $\min_{\lambda}\sum_{\vec\sigma}P_{\alpha}(\vec\sigma)^{1-\lambda}P_{\beta}(\vec\sigma)^{\lambda}$, where $P_{\alpha}$ and $P_{\beta}$ are the probability distributions of the D regions in fish $\alpha$ and $\beta$.
The thick red squares show the families.
{\bf B.} Mutual information between fish and sequence {\em vs.} the entropy of fish. Each point is a subgroup of all 13 fish (excluding fish F), color-coded by its size (from dark blue to red). Filled circles are averages over groups of each size. Upper Inset: comparison between mutual information estimated from counting observed sequences, and that predicted by the maximum entropy model. Lower Inset: Mutual information {\em vs.} fish entropy, as predicted by the independent model.
\label{panel4}}
\end{figure}

\subsection{Comparison between fish}

The analysis of entropies shows that the repertoires of individual fish span only a tiny fraction of the possible sequence space.   Do the repertoires of different fish overlap with each other, or are they distinct?   To answer this question, we first computed a similarity factor $\mathrm{Sim}[P_{\alpha},P_{\beta}]$ between repertoire distributions (see {\refappendix}). This factor takes values between 0 and 1 and measures the difficulty of guessing to which of the two repertoires ($\alpha$ or $\beta$) a given sequence belongs. Figure \ref{panel4}A shows the similarity factor for all pairs of fish, as calculated by the maximum entropy model (see {\refappendix}).
While the choices of V, D and J segments  are correlated with the family relations among the fish \cite{Weinstein:2009p1566}, this measure of similarity among D regions is not.

To study repertoire specificity beyond two fish, we looked at the mutual information between the identity $\alpha$ of a fish  and the sequence $\bs$ of a single antibody molecule, 
\begin{equation}
I(\alpha; \bs ) \equiv \sum_{\bs,\alpha} P(\bs,\alpha)\log_2\left[{{P(\bs,\alpha)}\over {P(\bs)P(\alpha)}}\right],
\end{equation}
where $P(\bs,\alpha)$ is the probability that a sequence picked at random in the dataset be $\bs$ and come from fish $\alpha$. Figure \ref{panel4}B represents this mutual information as a function of the fish entropy $S_{\alpha} = -\sum_\alpha P(\alpha)\log_2 [P(\alpha)]$ for many subgroups of fish of various sizes. The fish entropy is an upper bound to the mutual information, and is only reached when sequences give perfect information about which fish they came from, {\em i.e.} when each sequence belongs to one fish uniquely. Although the mutual information remains far from this upper bound,
it keeps growing linearly with the entropy as the size of the group is increased, each fish adding its own unique diversity; the slope of information vs. entropy is roughly $0.5$, so that half of the diversity is unique to each individual and half is shared across the population.   Importantly, this individuality of the sequence ensembles depends dominantly on correlations, since in the independent model, $P_{\rm ind}(\bs )$ from Eq (\ref{Pind}), the mutual information between identity and sequence is roughly a factor of four smaller (inset to Fig \ref{panel4}B).
All 13 fish do not suffice to cover the potential diversity of D regions, as evidenced by the absence of saturation.

\begin{figure}
\noindent\includegraphics[width=\linewidth]{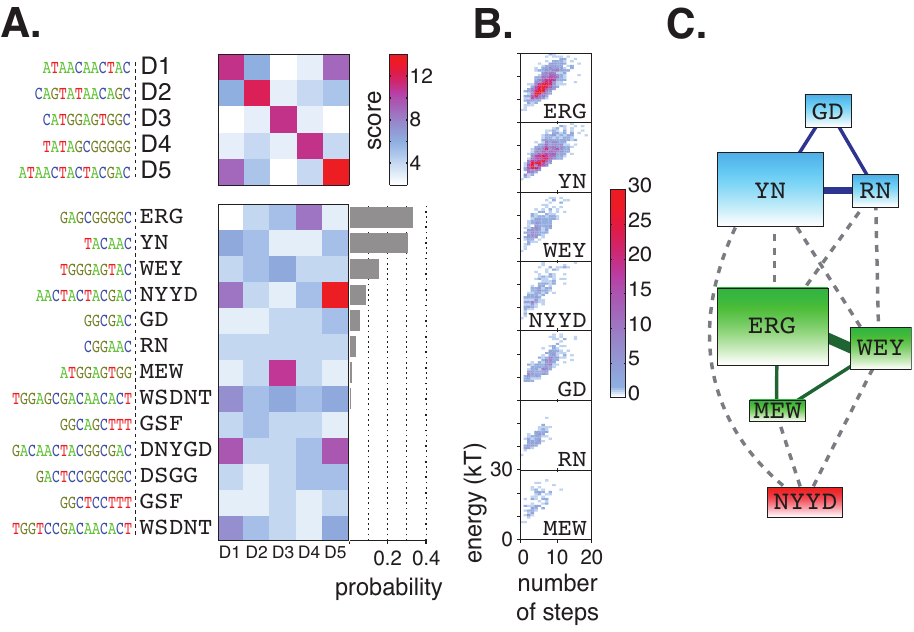}
\caption{{\bf Metastable states} (data from fish A). {\bf A} Lower: scores of pairwise alignments between the genomic segments D1-D5 and the metastable states. The bar plot represents the total weight of the basins of attraction of each metastable state. Upper: scores of alignments of the genomic segments with themselves and with each other are shown for comparison. {\bf B.} Basins of attractions of the seven most populated states. A density plot represents the energy of the sequences {\em vs.} the number of steps separating them from their metastable state by steepest descent. {\bf C.} Connectivity of the sequence space. Lines indicate the existence of paths of adjacent sequences between two metastable states. When the link is a solid line, there exists a path made only of single-nucleotide mutations.
\label{panel5}}
\end{figure}

\subsection{Multivalley landscape}

The energy function in Eq (\ref{energy}) includes competing interactions---the couplings $J$ can be positive or negative, favoring both correlated and anti--correlated amino acid substitutions at different sites.  From the statistical mechanics of disordered systems \cite{Mezard} we know that such competition can lead to `frustration' and many metastable states.
A metastable state is defined as a local mimimum of the energy landscape or, in probabilistic language, a local maximum of the probability distribution.   Does this happen in the case of antibody diversity?

Our model assigns an energy to every sequence, but to find local minima in this landscape we need to define `local.'   Since mutations occur at the level of nucleotides,  we work in the space of nucleotide sequences; to assign a (free) energy to nucleotide sequences we translate to amino acids, compute $E(\bs )$ from Eq (\ref{energy}), and add a correction term for the entropy of codon usage.  Then we say that two sequences are adjacent if: {\em (i)} they differ by one nucleotide; or {\em (ii)} they differ by one nucleotide insertion and one deletion; or {\em (iii)} they differ by three insertions or three deletions; the last criterion is necessary because, by construction, the lengths of D regions is a multiple of 3. With this conservative definition, we find $\sim10$ local minima per fish; examples are shown in Fig \ref{panel5}.  Some of these states correspond to the D regions encoded in the genome, as shown in Fig  \ref{panel5}A, but many do not.  The structure of the energy landscape, and hence the probability with which sequences appear in the organism's antibody repertoire, thus has elements which are not simply a record of genomic history, but presumably reflect rapid adaptation to the antigenic environment.

Each metastable state defines a basin of attraction or valley in the energy landscape, and we can assign each sequence to its corresponding valley
by moving `downhill:'   starting from a given sequence, go to the lowest energy neighbor, and continue doing so until the energy stops decreasing and a metastable state has been reached. 
Figure \ref{panel5}B represents the energy of all sequences in a basin of attraction as a function of their distance (in number of steps) to the metastable states; although there are differences of detail, the different basins have very similar structures.  As we explore away from the minimum energy in each basin, at some point we reach the `pass' which connects neighboring valleys; the trajectories over these passes are analogous to the trajectories from reactants to products in a chemical reaction, with the pass identifiable as the transition state \cite{Hanggi:1990p3202}.  Since the sequences are not too long, we can find these paths by a conventional Monte Carlo procedure (see {\refappendix}), and in most cases we found continuous paths through adjacent observed sequences between metastable states. When the two metastable states had the same length, we found paths where each step was a single nucleotide mutation.
Figure \ref{panel5}C summarizes the connections among the seven most populated metastable states in the repertoire of fish A.
Taken together, these  results on the energy landscape  imply that the repertoire explores much of the sequence space, and is not slaved to the genomic templates or to any specific sequence arising in the adaptation process.

\section{Summary and Discussion}

The formation of the antibody repertoire is an example of an accelerated evolutionary process under selective pressure. Antibodies in a given organism are correlated both through their genomic origin and as a result of the adaptation history. In this study we have analyzed the repertoire of B cell antibodies by building compact models of the hypervariable region of their heavy chain, based on the principle of maximum entropy.

The reduction of parameters achieved by the model is enormous.   Even though we are looking at the relatively short hypervariable D regions, there are tens of millions of possible sequences, and in principle each sequence occurs with a different probability in the repertoire.  
In constrast, the number of parameters of our model is of order $400K$, where $K$ is the interaction range. Importantly, this number scales reasonably with sequence size, making our approach tractable for systems in which the relevant sequence is much longer, including   the hypervariable regions in other species.
The compactness of the model allows for generalization, so we can predict quantities that are not deducible simply by counting sequences in the observed sample:  the overall size of the repertoire, the overlaps between repertoires of different individuals, and the probability of finding new, as yet unobserved, sequences in larger samples from the same individual.

The maximum entropy construction accounts for correlations between amino acid substitutions at different residue positions through an effective interaction structure. These interactions are strong enough to generate a dramatically different ensemble of sequences than would be expected if substitutions at each site were independent.  The diversity of the repertoire is substantially reduced (from an entropy of $\sim 14\,{\rm bits}$ to $\sim 8\,{\rm bits}$),
the distribution of sequences obeys Zipf's law, and the distribution has a complex structure of `metastable states,'  clusters of sequences with high probability. 

We have addressed the question of individuality, using our model and tools from information theory. At one extreme, the fish could be completely different, and each new fish would bring a whole new set of unique sequences. At the other extreme, fish could have more or less identical repertoires, sharing the same antibodies in the same proportions. We found an intermediate situation, where about $50\%$ of the repertoire diversity was unique to each fish, and the rest shared among all fish. As one concatenates the individual  repertoires, including more and more fish, the size of the resulting meta--repertoire must saturate, since the number of possible antibody sequences is finite. But this saturation is not reached even for thirteen fish, meaning that each fish is still unique compared to all other twelve taken together, and not only compared to each of them separately.

The details of the adaptation process undergone by the repertoire are largely unknown, and our model only provides a first step to aid in its study. What is the mutation mechanism? How do recognition and selection work? 
Our observation of Zipf's law provides an important constraint on these mechanisms.  As we have emphasized, this behavior arises only if the interactions between substitutions at different sites have a critical strength.
But these interactions are just a summary of the mutation and selection dynamics.  There are simple growth processes with mutation that can generate Zipfian distributions \cite{Yule:1925p3020}, but much work remains to find a realistic model that generates the full structure of $P(\bs )$.

The structure of the energy landscape underlying our model shows that the repertoire decomposes into several components. Each  component is centered on a metastable state,  a peak in the probability distribution of sequences. Some metastable states are closely related to the genomic templates, although rarely identical, while others are not attributable to any genomic template. We can think of these metastable states as markers of adaptation. For example, an infection could have caused the proliferation of  antibodies particularly efficient for recognizing a specific antigen, thus creating a peak in the probability landscape. This suggests the possibility of using metastable states and their basins of attraction for probing infectious history, perhaps in experiments that follow the dynamics of the sequence ensemble over time.

The clusters associated with the metastable states are not completely disconnected from one other: we found continuous paths of observed sequences between most metastable states. This means that, far from being slaved to their genome, the D sequences have the freedom to explore sequence space extensively during the adaptation process, forming a large cloud of possibilities between the higly concentrated regions of the sequence space, {\em i.e.} the metastable states, whether they be genomic or not. The method we have used for finding these paths---a Metropolis walk in energy space---further illustrates the power of the maximum entropy model: since it naturally favors low energy barriers, this algorithm is more likely to find paths where all sequences are present in the data. More generally, it could be used as a tool for retracing mutation paths between any two sequences, and could lend us insight into the repertoire's evolutionary history.

Finally, the success of maximum entropy models in accounting for the higher order statistical structure of the sequence ensemble encourages us to think that this approach is more widely applicable.   The maximum entropy formalism shows how, as in many statistical physics problems, the observable correlations between  amino acid substitutions at any two sites provide the signatures of collective behavior in the system as a whole.   The idea that crucial aspects of life should be viewed as emergent, collective phenomena has been discussed for decades.  The challenge has been to move beyond metaphor by developing precise mathematical tools for extracting quantitative models of this collective behavior from experiment. We believe that we have taken useful steps in this direction in the work reported here.

\begin{acknowledgments}
We thank SR Quake, JR Weinstein and their colleagues for sharing their data, and for several helpful discussions. This work was supported in part by NSF Grant PHY--0650617 and by NIH Grant P50 GM071598; T.~M. was supported by the Human Frontiers Science Program.
\end{acknowledgments}

\section*{APPENDIX}

\subsection{Aligning sequences} 

The dataset \cite{Weinstein:2009p1566} consists of a list of $\sim 200$ nucleotide long sequences for each fish. Each read is
aligned to a V and J genomic template using the Smith--Waterman algorithm \cite{Smith:1981p4415} with uniform scoring matrix (1 for a match and -1 for a mismatch), and then we isolate the subsequence that starts after the last nucleotide aligned to V, and ends before the first nucleotide aligned to J. This subsequence is then aligned with each of the five D templates, and assigned to the template that gave the best alignment score. To estimate the quality of this assignment, we calculate the score difference between the first and second best matches, divided by the same quantity assuming that the sequence is exactly identical to its genomic template. This ratio, which we call discriminability, is 0 when the two best genomic matches have the same score, and 1 when the sequence to assign is identical to a genomic template. Fig.~\ref{panel1}B shows the distribution of the discriminability ratio across all sequences in fish A.

Reads that have a stop codon, or for which the reading frames of the V and J segments are not congruent, are discarded. The remaining reads are translated into aminoacid sequences, which are aligned with their V and J genomic aminoacid sequences. Our analysis focuses on the D-region aminoacid subsequence, which is composed of the residues located strictly between the last aligned residue of V and the first aligned residue of J.
The length distribution of the obtained D regions is shown Fig.~\ref{panel1}A.

For the purpose of measuring statistical quantities, each sequence read $\bs$ is assigned a weight $w(\bs)$ inversely proportional to the PCR bias associated to its sequence primer \cite{Weinstein:2009p1566}. Thus the mean of any observable $\mathcal{O}(\bs)$ will be estimated using:
\beq\label{bias}
\<\mathcal{O}\>=\frac{\sum_s w(\bs)\mathcal{O}(\bs)}{\sum_s w(\bs)}.
\eeq

\subsection{Diversity across and within VDJ classes}

To estimate the variability and correlations among residues, we calculated the mututal information between pairs of residues at positions $i$ and $j$ of the sequence, which measures the degree of correlation between two positions, and is defined by:
\beq
I_{\rm ij}=\sum_{\sigma,\sigma'} P_{\rm ij}(\sigma ,\sigma')\log_2\frac{P_{\rm ij}(\sigma,\sigma ')}{P_{\rm i}(\sigma)P_{\rm j}(\sigma' )}
\eeq
where $P_{\rm i}(\sigma )$ is the probability of having residue $\sigma$ (=~Ala, Arg, Asn, $\ldots$) at position $\rm i$, and $P_{\rm ij}(\sigma ,\sigma' )$ the probability of having $\sigma$ at position $\rm i$ and $\sigma'$ at position $\rm j$. These probabilities are estimated by frequency count, weighted by the primer dependent PCR bias as in Eq (\ref{bias}).  For the purpose of comparing sequences at the same positions, we performed a multiple alignment of all the
sequences. For each read, the V and J regions were aligned to their
genomic template. Then, all 39 V and all 5 J genomic templates were
aligned with the multiple sequence alignment software ClustalW. Since
we could find no satisfactory multiple alignment of the D regions or
even the 5 D templates, we arbitrarily aligned all D regions by
their first amino acids.

The left panel of Fig.~\ref{panel1}C shows that antibody sequences indeed are highly variable and show large pairwise correlations, in agreement with the recombination scenario. Diversity in the choice of the V, D, J genomic segments induces variability at each residue position (as measured by the diagonal terms of the mutual information matrix, {\em i.e.} the single-position entropies), and these variables are themselves correlated through their common genomic cause (as measured by the off-diagonal terms).

We can estimate the part of diversity that is {\em not} due to recombination by calculating the average mutual information within VDJ classes:
$I_{\rm ij}^{\rm (cond)}=\< I_{\rm ij}(VDJ)\>_{VDJ}$ (where $I_{\rm ij}(VDJ)$ is the mutual information within VDJ),
shown in the right panel of Fig \ref{panel1}C.
This mutual information is small throughout the V and J segments, indicating that the choice of the V and J templates is the main source of diversity in these regions. In contrast, the D region remains diverse even within D classes.

\subsection{Solving the maximum entropy model}
For a set of parameters $(\mu,h,J)$, the observables $P^{\rm (m)} (L)$, $P^{\rm (m)}_1$, and $P^{\rm (m)}_2$ are estimated by the Metropolis algorithm. Starting from the independent model as initial condition: $\mu(L)=\log P(L)$, $h(\sigma)=\log P_1(\sigma)$, and $J_k=0$, we implement the following update rules:
\bea
\mu(L)&\leftarrow &\mu(L)+\epsilon_1 \log\frac{ P(L)}{P^{\rm (m)}(L)},\\
h(\sigma)&\leftarrow &h(\sigma)+\epsilon_1 \log\frac{ P_{1}(\sigma)}{P^{\rm (m)}_{1}(\sigma)},\\
J_{\rm k}(\sigma,\tau)&\leftarrow &J_{\rm k}(\sigma,\tau)+\epsilon_2 \log\frac{ P_{2}(\sigma,\tau; k)}{P^{\rm (m)}_{2}(\sigma,\tau; k)}.
\eea
The last of these three rules is implemented every 5 steps, while the first two rules are implemented the remaining four steps. We set $\epsilon_1=0.005$ and $\epsilon_2=0.01$.

\begin{widetext}
\subsection{Entropy calculations}

To calculate the entropy of the probability distribution $P$, we need an estimate of the number $P(\bs)$ for each D region sequence $\bs$. We estimate this number using Eq.~\ref{bias}. The distribution is undersampled due to the large number of possible sequences compared to the actual number of reads, and therefore we expect some systematic error. To reduce that error we compute the entropy with the method described in \cite{Strong}, which extrapolates the infinite-sample limit from finite-sample estimates.

The entropy of the model distribution $P^{\rm (m)}$ can, on the other hand, be calculated with arbitrary precision using thermodynamic integration. We define
\beq\label{Zbeta}
Z(\beta)=\sum_{\bs} \exp\left[\mu(L)+\sum_{{\rm i}=1}^L h(\sigma_{\rm i})
+\beta \sum_{{\rm k}=1}^{K} \sum_{\substack{{\rm i,j}\\{\rm i}-{\rm j}={\rm k}}} J_{\rm k}(\sigma_{\rm i}, \sigma_{\rm j})\right],
\eeq
\end{widetext}
such that $Z(1)=Z$, and $Z(0)=\sum_L e^{\mu(L)} {\left[\sum_{\sigma} e^{h(\sigma)}\right]}^L$. We obtain $Z$ by calculating the following integral numerically:
\beq
\log Z(1)=\log Z(0) +\int_{0}^1 d\beta \, \frac{d\log Z(\beta)}{d\beta},
\eeq
with
\beq
\frac{d\log Z(\beta)}{d\beta}=\sum_{\substack{{\rm i,j}\\{\rm i}-{\rm j}={\rm k}}} \left\<J_{\rm k}(\sigma_{\rm i}, \sigma_{\rm j})\right\>_{\beta},
\eeq
where the mean $\<\cdot\>_{\beta}$ is taken with weights given by Eq (\ref{Zbeta}). This quantity can be computed by the Metropolis algorithm for each $\beta$. Finally, the entropy is given by:
\beq
S[P^{\rm (m)}]=\log Z + \< E(\bs) \>_{\beta=1},
\eeq
where the second term is also computed by Metropolis.

\subsection{Zipf's law and criticality}

We are describing the distribution of states $\bs$ in the Boltzmann form, Eq (\ref{boltzmann}).  Then a natural quantity is the density of states, 
\begin{equation}
\rho(E) = \sum_{\bs}\delta\left[E-E({\bs})\right] .
\end{equation}
In the limit of a large system, $\rho(E)$ becomes smooth, so that   $n(E) = \rho(E)\Delta$ is the number of states with energy $E$, assuming we have some finite resolution $\Delta$.  The log of this number is the entropy, $\ln n(E) = S(E)$.
Further, in the thermodynamic limit we expect that both entropy and energy are extensive variables, so we can define the energy and entropy per degree of freedom,\begin{eqnarray}
\epsilon &=& E/N\\
s(\epsilon ) &=& S(E = N\epsilon) / N ,
\end{eqnarray}
where $N$ is the number of degrees of freedom.  
It's hard to ``measure'' the density of states $\rho(E)$, but it's easier to think about the number of states with energy less than $E$, which we'll call ${\cal N}(E)$.  By definition,
\begin{equation}
{\cal N}(E) \equiv  \int^E dE' \, \rho(E' ).
\end{equation}
At large $N$, we then have 
\begin{eqnarray}
{\cal N}(E) &=& {1\over \Delta}  \int^E dE' \,\exp[S(E)]\\
&=& {N\over \Delta} \int^{E/N} d\epsilon' \, \exp[Ns(\epsilon')]\\
&\approx& {N\over \Delta} e^{Ns(E/N)} \int_0^\infty dx\, e^{-Ns'(E/N) x}
\\
&=&{1\over{\Delta s'(E/N)}}e^{Ns(E/N)} .
\end{eqnarray}
But the rank of state $\bs$ is exactly the number of states with energy smaller that $E({\bs})$, 
\begin{equation}
r_{\bs} = {\cal N}\left[E = E({\bs}) \right] .
\end{equation}
With our expression for $\cal N$ we can write
\begin{equation}
\ln (r_{\bs} ) = N s\left[E({\bs})/N\right] - \ln\left[ \Delta s'(E/N)\right] ,
\end{equation}
which is dominated by the first term at large $N$.  Then Zipf's law, $P(\bs ) =  A/r_{\bs}$ means that
\begin{equation}
\ln P(\bs ) = - N s\left[E({\bs})/N\right] + \ln\left[ A\Delta s'(E/N)\right] .
\end{equation}
But from the Boltzmann distribution we have
\begin{equation}
\ln P(\bs ) = -  E({\bs}) - \ln Z ,
\end{equation}
so Zipf's law implies
\begin{equation}
N s(E/N) = S(E) = E + \cdots ,
\end{equation}
where $\cdots$ denotes terms independent of $E$ or terms which vanish relative to $S$ in the large $N$ limit.

\bigskip

\subsection{Similarity factor}

 The similarily factor between two  distributions $P_\alpha$ and $P_\beta$ is defined as $\mathrm{Sim}[P_{\alpha},P_{\beta}]= \min_{\lambda}\sum_{\bs}P_{\alpha}(\bs)^{1-\lambda}P_{\beta}(\bs)^{\lambda}$. The similarity factor appears naturally in the asymptotic error of the following discrimination task. Suppose that we know $P_{\alpha}$ and $P_{\beta}$, which describe the repertoires of two fish $\alpha$ and $\beta$, and that we are given $N$ sequences $\bs^1,\ldots,\bs^N$, which either all come from fish $\alpha$, or from fish $\beta$. What is the probability of attributing the sequences to the wrong fish, and how does it decay with the number of observations $N$?
 
The log--likelihood for $N$ sequences $\{\bs^1,\ldots,\bs^N\}$ coming from $\alpha$ is $\sum_{i=1}^N \log P_\alpha(\bs^i)$, and likewise for $\beta$.
One infers that the sequences came from $\beta$ if the log--likelihood for $\beta$ is larger, and vice--versa. Thus the probability of error is, assuming the sequences came from $\alpha$:
\beq
\mathbb{P}({\rm error}) =\sum_{\bs^1,\ldots,\bs^N} \prod_{i=1}^N P_{\alpha}(\bs^i)\theta\left[\sum_{i=1}^N\log \frac{P_{\beta}(\bs^i)}{P_{\alpha}(\bs^i)}\right].
\eeq
Using the integral representation of $\theta(x)$ and a saddle-point estimate, we find the asymptotic behavior of the error probability for large $N$:
\beq
\mathbb{P}({\rm error}) \asymp \left[\min_{\lambda}\sum_{\bs}P_{\alpha}(\bs)^{1-\lambda}P_{\beta}(\bs)^{\lambda}\right]^N,
\eeq
which is symmetric in $\alpha$ and $\beta$, and is exactly $\left(\mathrm{Sim}[P_{\alpha},P_{\beta}]\right)^N$.

For each $\lambda$, the sum $\sum_{\bs}P_{\alpha}(\bs)^{1-\lambda}P_{\beta}(\bs)^{\lambda}$ can be computed directly from frequency counts, or, in the case of the model distribution, by thermodynamic integration as explained above. The minimum over $\lambda$ is obtained by a line minimization search.

\end{document}